\documentclass{iau}
\usepackage{graphicx}

\title[] %% give here short title %%
{The gravitational wave signal from isolated objects}

\author[Jinzhong Liu \& Yu Zhang]   %% give here short author list %%
{Jinzhong Liu $^1$
 \and  Yu Zhang $^1$}

\affiliation{$^1$National Astronomical Observatory/Xinjiang
Observatory, Chinese Academy of Sciences, 150 Science 1-Street,
Urumqi, Xinjiang 830011, China  \\ email: {\tt liujinzh@xao.ac.cn } \\[\affilskip]
}

\pubyear{2012}
\volume{290}  %% insert here IAU Symposium No.
\jname{Feeding compact objects: Accretion on all scales}
\editors{C.M. Zhang, T. Belloni, M. M\'endez \& S.N. Zhang, eds.}

\begin{document}

\maketitle

\begin{abstract}
According to the theoretical study, a deformation object (e.g., a
spinning non-axisymmetric pulsar star) will radiate a gravitational
wave (GW) signal during an accelaration motion process by LIGO
science project. These types of disturbance sources with a large
bump or dimple on the equator would survive and be identifiable as
GW sources. In this work, we aim to provide a method for exploring
GW radiation from isolated neutron stars (NSs) with deformation
state using some observational results, which can be confirmed by
the next LIGO project. Combination with the properties in
observation results (e.g., PSR J1748-2446, PSR 1828-11 and Cygnus
X-1), based on a binary population synthesis (BPS) approach we give
a numerical GW radiation under the assumption that NS should have
non-axisymmetric and give the results of energy spectrum. We find
that the GW luminosity of $L_{GW}$ can be changed from about
$10^{40}\rm erg/s$-- $10^{55}\rm erg/s$.

\end{abstract}

\firstsection % if your document starts with a section,
\section{Introduction}

In Einstein¡¯s theory of general relativity, gravitational wave (GW)
is considered as a phenomenon resulting from a space perturbation of
the metric traveling at the speed of light, and the observation of
the binary pulsar  PSR 1913+16 has given an indirect evidence of GW
radiation (e.g. Hulse \&Taylor 1975). Nowadays, these ripples in
space-time due to GW have still not been directly observed on the
ground detectors. The various frequency ranges of the GW detectors
can respectively fix different GW sources (Jaffe \& Backer 2003;
Belczynski, Kalogera \& Bulik 2002; Liu 2009; Liu et al. 2010A; Liu
et al. 2010B; Liu et al. 2012). Here, we focus on the other, much
less studied groups of isolated neutron stars due to asymmetric
mechanism, which can be divided into two groups according to the
difference of GW radiation: I) the intrinsic asymmetry of NSs, II)
the relative motion of the asymmetric part of NSs (e.g., Papaloizou
\& Pringle 1978 ). In this work, we aim to explore the GW radiation
from group I. The three formation mechanisms in group I can be
summarized as follows: i) a rotating NS with asymmetrical ellipsoid.
(e.g., Hessels et al. 2006); ii) a oblique-dipole-rotator model
(e.g., PRS1828-11); iii) the mass deformations due to an
eccentricity in the equatorial plane of NS (the typical example is
the low-mass X ray binaries: Cygnus X-1). The purpose of this poster
is to study the GW radiation from an isolated object with asymmetric
structure.

\section{Computations}

The GW luminosity $L_{GW}$ and dimensionless strain h of a rigidity
object with rotating process are predicted by Press \& Thore (1972).
The description of our physics parameters and assumptions are as
follows: (I) In the single-star evolution code, we trace back to the
formation of a NS from the zero-age main sequence to remnant stages.
(II) We give the fitting curves of physical properties  according to
the NS dynamic structure model and equation of state in left panel
of Fig. 1.(III) For the eccentricity e, we obtain it from uniform
distribution in the range $10^{-3}$ to $10^{-11}$.(IV)We present a
Gaussian distribution of D from the GW sources to the earth in the
Galaxy. In order to compare with observation, we download 109 NSs
with rotation period less than 0.05s from the website
(http://www.atnf.csiro.au/research/pulsar/psrcat/). In middle panel
of Fig. 1, we give the distribution of the rotation period between
our model results and observations.

\begin{figure}[!htp]
% \vspace*{-2.0 cm}
\begin{center}
 \includegraphics[height=1.2in,width=1.6in]{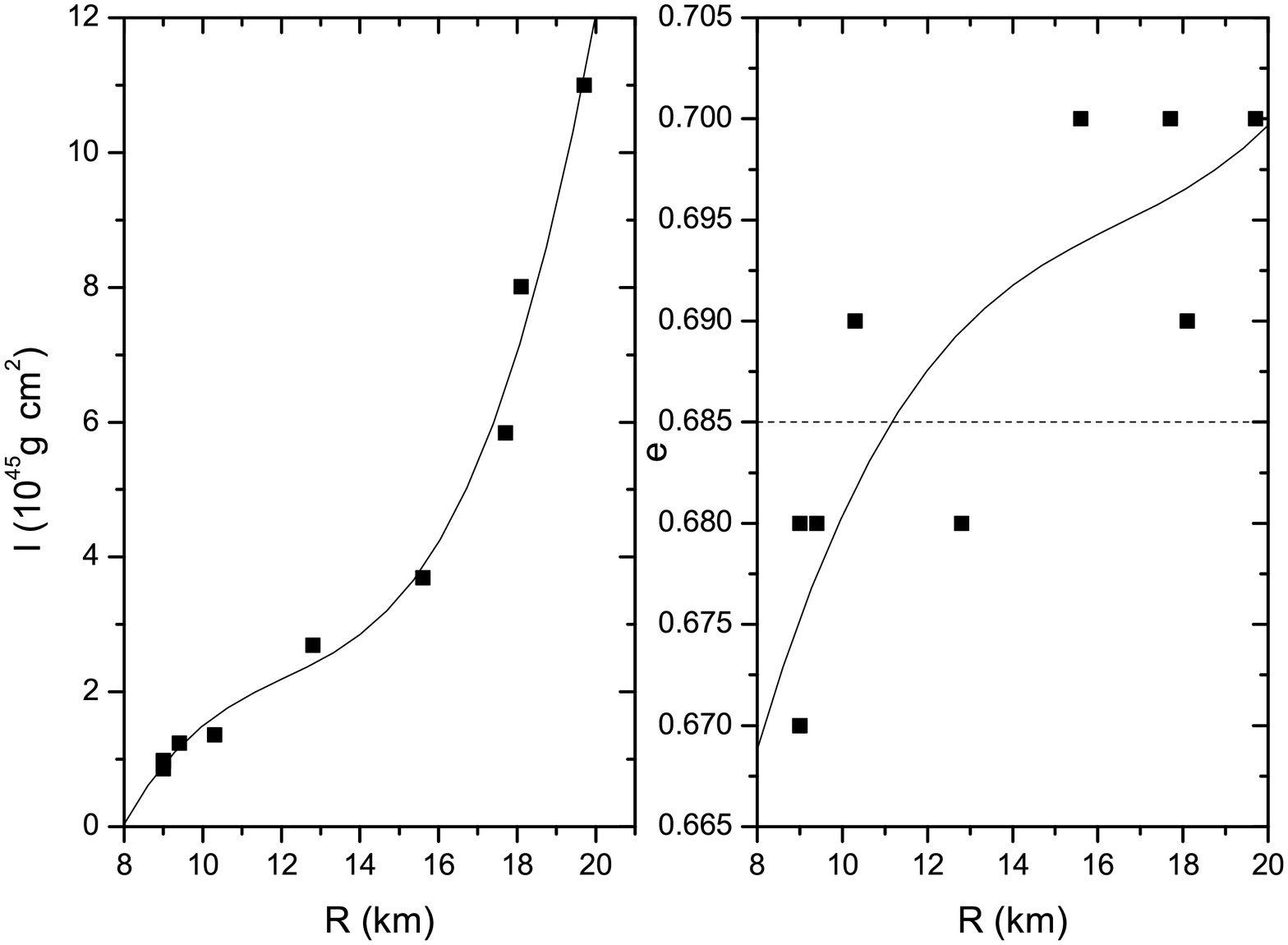}
%  \includegraphics[height=1.2in,width=1.6in]{liu10.eps}
%\includegraphics[height=1.2in,width=1.6in]{liu11.eps}
% \vspace*{-1.0 cm}
 \caption{Left:The fitting physical property curves of NS; Middle:The distribution of  rotation period; Right:The spectral energy distribution of rapid rotating NS. }
   \label{fig1}
\end{center}
\end{figure}

\section{Results and Discussion}
In general, the total energy ($Mc^{2}$) of NS is about $10^{54}$erg,
the rotation energy is $~ 10^{ 53}$ erg. Therefore, for the
e$>10^{-5}$, the most energy of NS can be radiation as GW signal
during several years. All these calculations of isolated objects can
examine that these sources are expected to be a type of GW sources.
In our model, the influence parameter $\xi$ of eccentricity can be
changed to the influence of mass, which is $ ~ 2.8\times10^{-4}<\xi<
8.9\times10^{-9}$, corresponding to the value of strain amplitude
$0.8\times10^{ -24} < h < 10^{ -32}$. Our prediction agrees with
that of Crab nebula and Virgo cluster ($10^{-24}-10^{-25}$).
Meanwhile, the right panel of Fig. 1 gives the spectral energy
distribution of rapid rotating NS and implies that the GW luminosity
of $L_{GW}$ can be changed from about $10^{40}\rm erg/s$--
$10^{55}\rm erg/s$.

\begin{acknowledgements}

 This work is supported by the program of the light in China's
Western Region (LCWR) (No. XBBS201022), Natural Science Foundation
(No. 11103054) and Xinjiang Natural Science Foundation (No.
2011211A104). This project/publication was made through the support
of a grant from the John Templeton Foundation and National
Astronomical Observatories, Chinese Academy of Sciences (No.
100020101).
\end{acknowledgements}

\end{document}